\documentclass[9pt,twocolumn,twoside]{optica}
\usepackage{hyperref}
\newcommand{\dois}[2]{\href{http://dx.doi.org/#1}{#2}}
\setboolean{shortarticle}{false}
\setboolean{minireview}{false}

\title{Talbot-enhanced, maximum-visibility imaging of condensate interference} 
\author{Y.\ Zhai}
\author{C.\ H.\ Carson} 
\author{V.\ A.\ Henderson} 
\author{P.\ F.\ Griffin}
\author{E.\ Riis} 
\author[1*]{A.~S.\ Arnold}
\affil{Department of Physics, SUPA, University of Strathclyde, Glasgow G4 0NG, United Kingdom}
\affil[*]{aidan.arnold@strath.ac.uk}

\dates{Compiled \today}

\ociscodes{(020.1475) Bose-Einstein condensates; (020.1335) Atom optics; (110.6760) Talbot and self-imaging effects} 

\doi{\url{http://dx.doi.org/10.1364/optica.XX.XXXXXX}}

\begin{abstract}
Nearly two centuries ago Talbot first observed the fascinating effect whereby light propagating through a periodic structure generates a `carpet' of image revivals in the near field. Here we report the first observation of the spatial Talbot effect for light interacting with periodic Bose-Einstein condensate interference fringes. The Talbot effect can lead to dramatic loss of fringe visibility in images, degrading precision interferometry, however we demonstrate how the effect can also be used as a tool to enhance visibility, as well as extend the useful focal range of matter wave detection systems by orders of magnitude. We show that negative optical densities arise from matter-wave induced lensing of detuned imaging light -- yielding Talbot-enhanced single-shot interference visibility of $>135\%$ compared to the ideal visibility for resonant light.  
\end{abstract}

\setboolean{displaycopyright}{false}

\begin{document}

\maketitle
\thispagestyle{fancy}
\ifthenelse{\boolean{shortarticle}}{\abscontent}{}

\section{introduction}

\begin{figure}[!t]
\center
\includegraphics[width=\columnwidth]{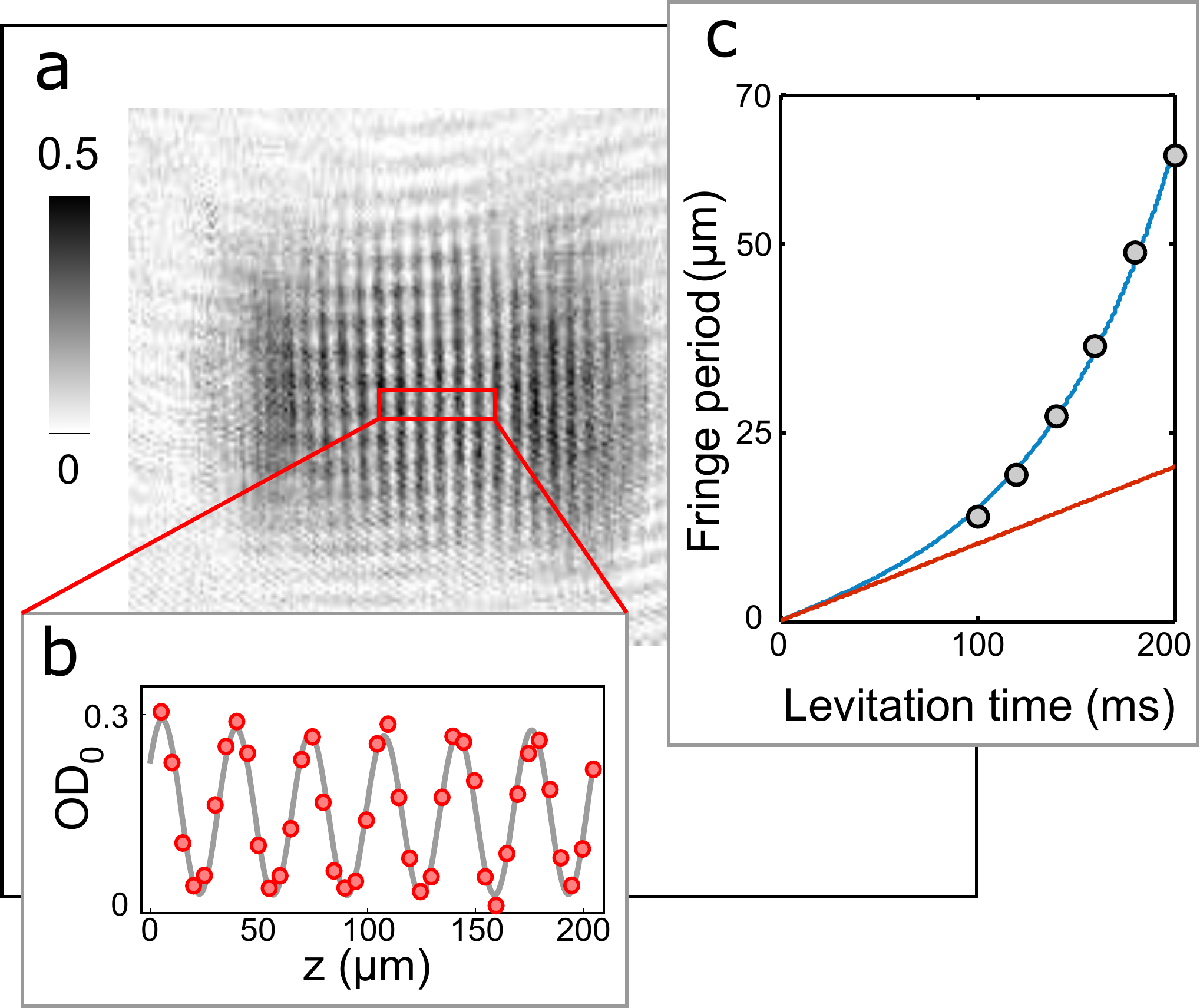}
\caption{Matter wave interference. Experimental absorption image (a, $1.4\times1.4\,$mm$^{2}$) of a condensate axially split by a blue-detuned optical dipole beam at $160\,$ms levitation time, scalebar indicates optical density (OD). Background-subtracted, angle-corrected, row-averaged OD profile (b, red points) of the red rectangular region in (a). The fitted sine wave has a parabolic spatial envelope and offset (gray curve) for fringe visibility and period extraction. Fringe period as a function of levitation time (c):  experimental data (black dots); inferred fringe period evolution (blue curve, Eq.~\ref{eq:newfringeperiod} with $\omega_z=13.8\pm 1.4\,$rad/s and $d=45\,\mu$m); ballistic expansion theory (red line, Eq.~\ref{eq:fringeperiod}).}
\label{fig:fig1}
\end{figure}

Since the first experimental realisation of gaseous Bose-Einstein condensates (BECs) \cite{FirstBEC,FirstBEC2} the field has grown dramatically and condensates are now used for a variety of studies including the flow of superfluids \cite{superfluid}, studies of strongly dipolar systems \cite{dipolar1, dipolar2}, 
and for producing atom-lasers \cite{BEClaser}. Matter-waves are also ideal for quantum technologies, a key ingredient of which is atom interferometry. Interferometers yield precise measurements of fundamental constants \cite{fundconstant}, rotation \cite{rotations,rotations2}, acceleration \cite {accelerations}, gravity \cite{gravity2,microgravity} and gravitational shifts \cite{gravity}. 
Interferometry is also a key method for studying the superfluid states prevalent in quantum degenerate systems \cite{super1,super2,super3,grimm,super4,super5}.

BECs are fully coherent matter wave sources, as first strikingly demonstrated in Ref.~\cite{Interference1}. A cigar-shaped single-well potential was transformed into a double-well by axial separation with an optical dipole beam. Ballistic expansion led to wavepacket spatial interference, reminiscent of Young's double slits. We use a similar configuration \cite{freespace}, but a simpler levitation-enhanced imaging technique, achieving improved visibility of $80\%$ (Fig.~\ref{fig:fig1}) above our previous record of $60\%$ by longer levitation and a weaker dipole beam. BEC interference has also been demonstrated via radial  \cite{Interference2,malcolm,atomchip,Interference3,folman} or axial \cite{Bragg,Standingwave} separation using a wide variety of other techniques. 

During our BEC interference investigations we observed that the fringe visibility strongly depended on the camera focal location, which we can now clearly attribute to the Talbot effect.  Talbot\rq{}s 1836 observation \cite{Talbot} of self-imaging of a periodic structure in near-field diffraction arises due to wave optics \cite{Rayleigh}. The phenomenon has since been studied in a wide variety of scientific disciplines including  optics, acoustics, electron microscopy, x-rays and plasmonics \cite{TalbotStudies,TalbotReview}. Transversely coherent waves with wavelength $\lambda$ passing through a regular phase/intensity structure with period $\lambda_f$ produce a self-image with a phase of $\pi$ $(0)$ at odd (even) integer multiples of the Talbot distance $\Lambda_{\textrm{T}}={\lambda_f}^{2}/\lambda$ \cite{TalbotDistance}. Fractional and fractal Talbot effects arise if the grating structure contains higher spatial harmonics \cite{berry}.

With matter waves the spatial Talbot effect has been observed from hot atomic beams \cite{TalbotMatterWave,TalbotMatterWave2} and the temporal Talbot effect has been seen with BECs in 1D optical lattices \cite{TalbotTimeDomain,TalbotTimeDomain2}. The effect is also an important trigger for spontaneous spatial light/matter pattern formation in cold atomic gases \cite{PatternInformation}. Although the regular period and phase of BEC interference patterns are extremely important for metrology, the dramatic influence of the Talbot effect on the propagation of near-resonant light used for imaging condensates has not been previously seen. 

Here we present the first observation of spatially periodic visibility variation in a BEC interferometer due to the Talbot effect. Imaging light is absorbed and refracted by $N$ BEC interference fringes, then propagates through the imaging system. At integer Talbot distances $i \Lambda_{\textrm{T}}$ ($|i|\lesssim N$) from the central image plane the diffracted light dramatically re-images to $N-|i|$ fringes  (Fig.~\ref{fig:fig2}). Furthermore, by detuning the imaging light, negative optical densities \cite{bosh} arise leading to enhanced visibility, nearly doubling the resonant value. 

Our results will be of particular relevance to atom interferometry experiments with several fringes ($N>5$), as opposed to those with only a few fringes or an overall population amplitude variation. Multi-fringe interferometers may have been optimised in the past based on high visibility, finding only a local maximum. In situations where the fringe period is critical, an assumption of only one image plane could lead to erroneous results (cf.~Fig.~2e). In the limit of short period fringes, with period around one wavelength, the Talbot distance of order $1\,\mu$m would lead to extreme difficulty to position the camera focal plane, with corresponding sensitivity to drift. This would be particularly important if e.g.\ single-atom absorption imaging \cite{kiel} were used in a quantum gas microscope \cite{greiner,kuhr}.



\section{experimental setup and fringes}

In our experimental setup \cite{freespace,setup}, a few $10^{8}$ $^{87}$Rb atoms in the weak-field-seeking state $|F=2,m_{F}=2\rangle$ are trapped at the top of a magnetic storage ring \cite{setup}. The Ioffe-Pritchard (IP) trap has axial and radial trapping frequencies of $10\,$Hz and $90\,$Hz, respectively, and a lifetime of $55\,$s. After $30\,$s RF evaporation, condensates with $5 \times 10^5$ atoms are created and observed using standard resonant absorption imaging on the stretched $D_2$ transition $|F=2,m_F=2\rangle \rightarrow |F'=3,m_F=3\rangle$, with magnification $\times 2$. A $658\,$nm blue-detuned repulsive dipole beam is used for axial BEC splitting with $<1\,$mW derived from a diode laser with an acousto-optic modulator focused to a $10\,\mu$m waist \cite{freespace}. 

Split BECs are released by rapidly switching off the optical and magnetic potential and they expand in space for sufficiently long that the final atomic cloud size is much larger than the initial distribution. Thus even initial spatial distributions with some broken symmetry behave like two matter wave point sources. This leads to clear `Young's-slit' type planar interference patterns in 3D, 
(Fig.~\ref{fig:fig1}a,b), obviating the need for specialised tomographic imaging \cite{Interference1}. 

The BEC fringe period $\lambda_f$ comes from the de Broglie wavelength of the two condensates and can be expressed as,
\begin{equation}
\lambda_f=\frac{h t}{md}
\label{eq:fringeperiod}
\end{equation}
where $h$ is Planck's constant, $t$ the ballistic expansion time, $m$ the atomic mass and $d$ is the initial BEC centre-of-mass separation. The free fall time is limited to $0-30\,$ms by the size of the imaging area, and times $<60\,$ms by the physical extent of the vacuum cell. To overcome these limitations we use a magnetic levitation field \cite{FirstBEC,Levitation,Levitation2} derived from a toroidal quadrupole \cite{setup} offset by an additional vertical constant field. BECs therefore experience a weak inverted parabola potential ($U_{z}=-m  {\omega_z}^{2}z^{2}/2$ with $\omega_z$ the angular velocity along the $z$ direction i.e.\ the BEC axis) due to the circular nature of the levitation coils, which we showed in Ref.~\cite{freespace} modifies $\lambda_f$ in Eq.~\ref{eq:fringeperiod} to:
\begin{equation}
{\lambda_f}'=\lambda_f\frac{\sinh({\omega_z} t)}{{\omega_z} t}.
\label{eq:newfringeperiod}
\end{equation}

Although our initial BEC separation $d$ is large in comparison to other experiments, the levitation potential we apply magnifies the spatial scale of our fringes \cite{freespace,grimm} (Fig.~\ref{fig:fig1}c). To achieve the equivalent maximum fringe period by ballistic expansion we would need $800\,$ms i.e.\ $3\,$m of free-fall. To resolve the fringes levitation times $\geq 70\,$ms are needed until the period corresponds to two or more camera pixels ($\geq 10\,\mu$m). 

Using resonant light absorption imaging fringe periods of $37.5(0.8)\,\mu$m are observed, with the highest visibility seen in this type of interferometer - 80\% (Fig.~\ref{fig:fig1}a,b) - close to the theoretical limit due to camera pixellation of the fringes. Depending on the timing and power of the dipole beam application either two phase-separate condensates are created or a single condensate can be split into two phase-coherently \cite{relative,Interference2}. The size of our interference pattern in comparison to our centre-of-mass uncertainty currently prevents us from ascertaining whether the relative phases of our split condensates are random or correlated, however the main results are unaffected.

\section{talbot effect: theory and experiment}

The effect of imaging light interacting with periodic BEC fringes can be modelled by Fourier-propagating the light waves, based on the initial conditions of the field immediately after the condensate, then performing the inverse Fourier transform to convert back to real space \cite{Fourier,Fourierb,Fourier2,Fourier3}. Propagation effects inside the BEC are neglected as the condensate size in the beam propagation direction is considerably smaller than the Talbot period. 

\begin{figure*}[!t]
\center
\begin{minipage}{.68\columnwidth}
\includegraphics[width=.99\columnwidth]{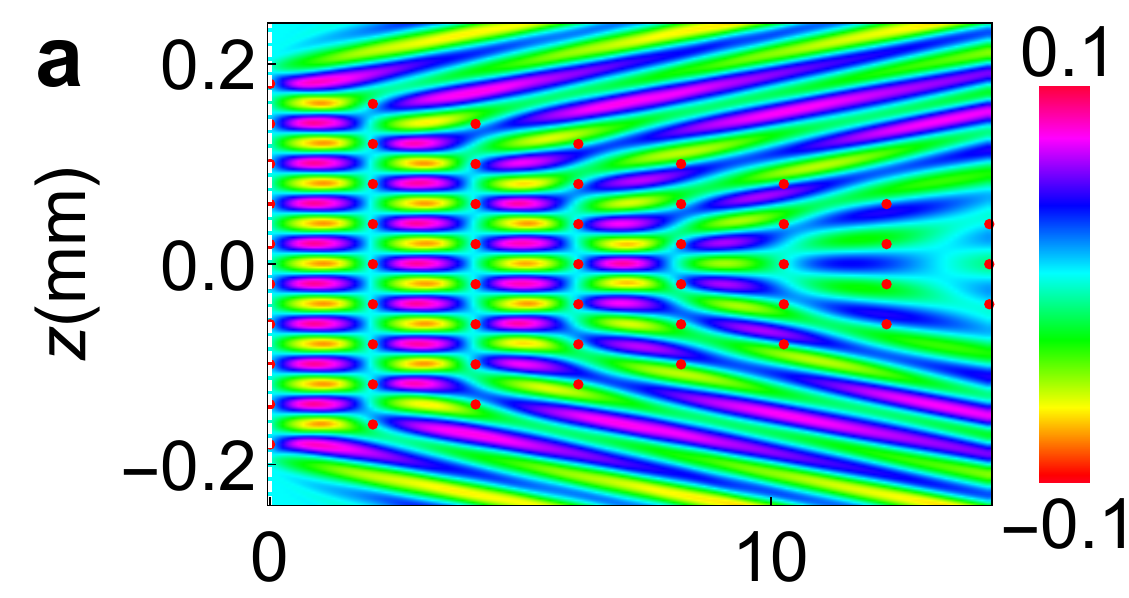}
\includegraphics[width=.99\columnwidth]{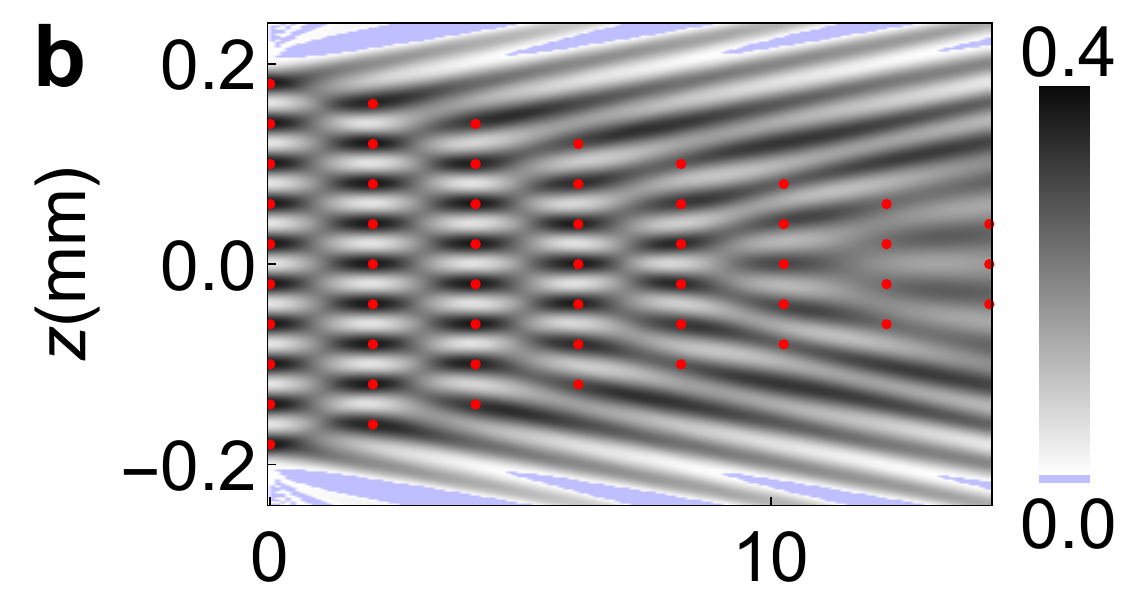}
\includegraphics[width=.99\columnwidth]{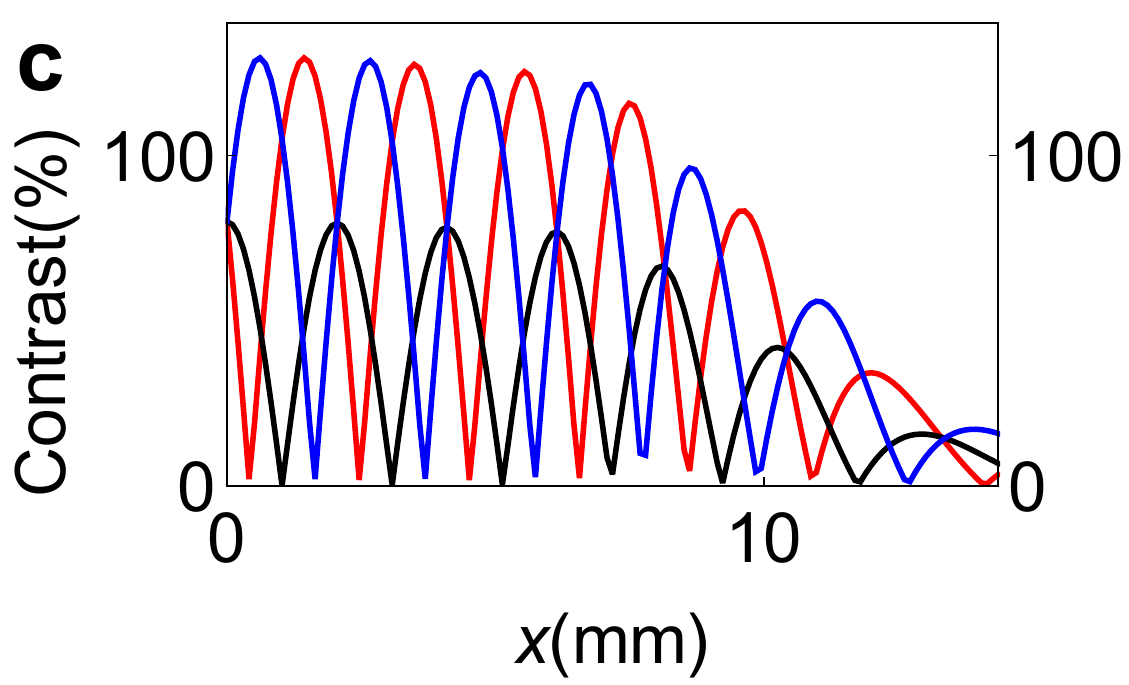}
\end{minipage}
\begin{minipage}{1.32\columnwidth}
\includegraphics[width=.99\columnwidth]{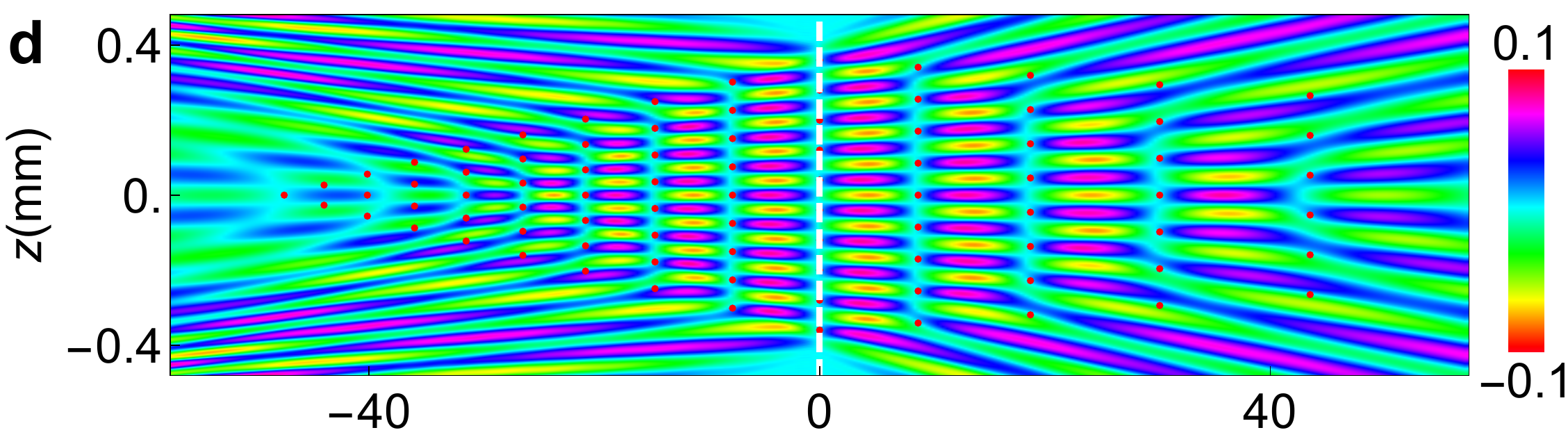}
\includegraphics[width=.99\columnwidth]{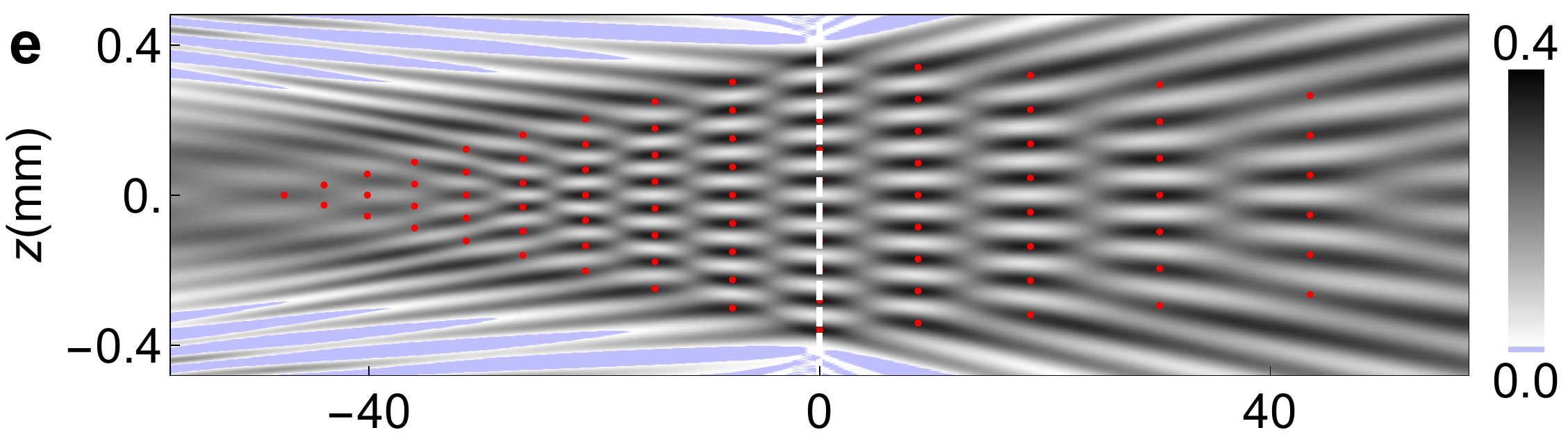}
\includegraphics[width=.98\columnwidth]{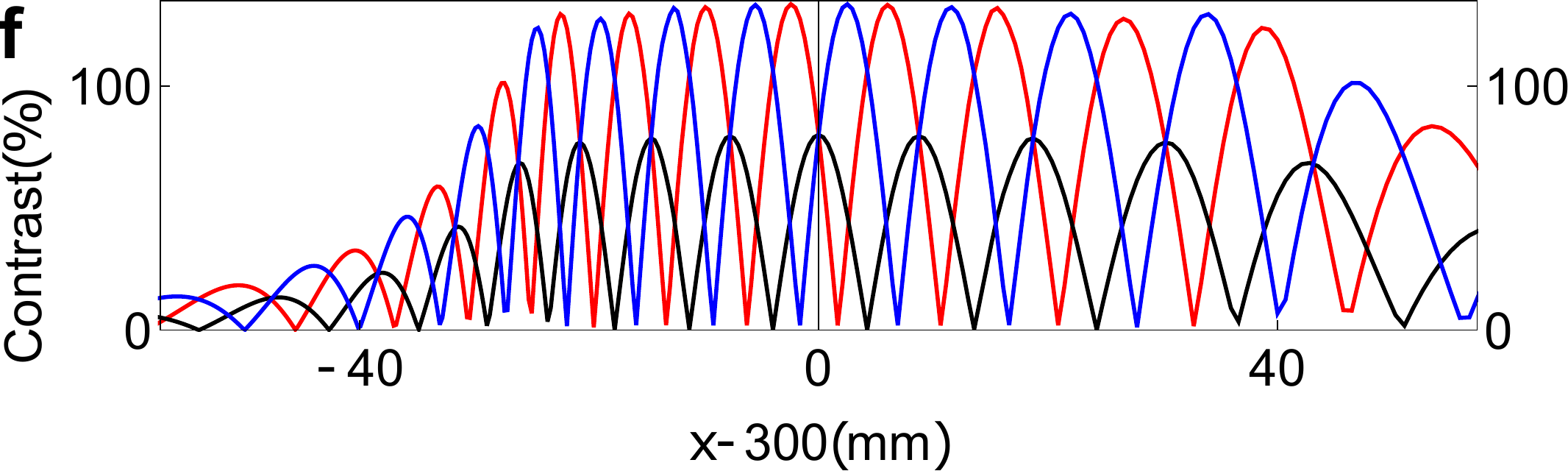}
\end{minipage}
\caption{Theoretical light propagation through BEC fringes. \textbf{a-c} Ten BEC fringes along $x=0$ of peak optical density OD$_0=0.4$ and visibility $80\%$ interact with resonant light leading, on propagation $(x>0)$, to periodic phase (\textbf{a} in radians) and OD (\textbf{b}) profile revivals. Antinodes of OD (red points) correspond to nodes of the phase, and vice versa. The central fringe at $z=0$ is displayed (\textbf{c}) for red, resonant (black) and blue laser detunings of -$\Gamma/2$, 0, and $\Gamma/2$, illustrating how detuning not only phase-shifts the fringes in $x$, but can enhance visibility relative to the initial optical density. After the light in images \textbf{a}, \textbf{b} and \textbf{c} propagates through a $300\,$mm long $\times 2$ magnification imaging system, the corresponding phase, OD and visibility is shown in \textbf{d}, \textbf{e} and \textbf{f} respectively. Note the chirp in the spatial period about the image plane (white dashed line in \textbf{d,e}).} 
\label{fig:fig2}
\end{figure*}

\begin{figure}[!t]
\center
\includegraphics[width=0.99\columnwidth]{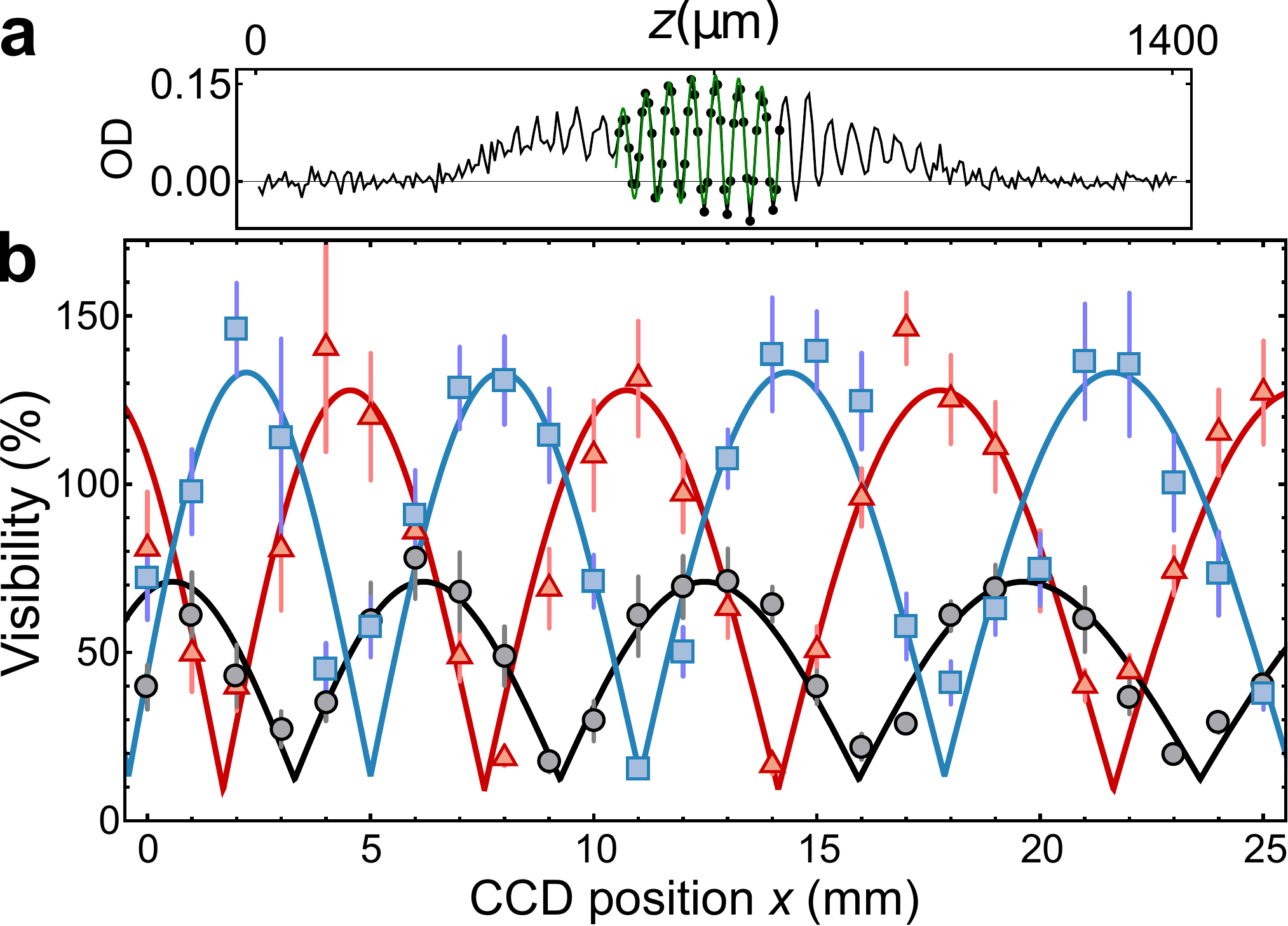}
\caption{\textbf{a} Single-shot 1D BEC OD profile (21 row average, $\Delta=+4\,$MHz), with visibility $C=1.5$ from fitting data points (black) with Eq.~\ref{fitt} (green curve). \textbf{b} Experimental (dots) and theoretical (curves from Fig.~\ref{fig:fig2}e) visibility as a function of camera position for detunings $-4, 0, +4\,$MHz (red, black and blue, respectively). All error bars in this paper represent the standard deviation of results from three separate images. 
The central Talbot period was fit to be $6.1(0.1)\,$mm, which agrees well with the theoretical Talbot period of $6.4(0.3)\,$mm predicted from the fringe period and Eq.~\ref{Eq:TalbotPeriod}.}
\label{fig:fig3}
\end{figure}

When light passes through the BEC fringes, the light attenuation and phase shift depend on the light detuning and the integrated atomic density. In the standard absorption imaging we use, the light passing through the atomic cloud indicates a transversely spatially dependent optical density 
\begin{equation}
\textrm{OD}=\ln\left(\frac{I_{i}-I_b}{I_{e}-I_b}\right), \label{ODeq}\end{equation}
where $I_{i}$, $I_{e}$ and $I_{b}$ are the light intensity distributions incident on the atoms, exiting the atoms, and due to background light, respectively. 
For low intensity light the optical density in the BEC pattern varies with detuning as $\textrm{OD}=\textrm{OD}_0/(1+4(\Delta/\Gamma)^2)$, where $\Delta$ is the light detuning from the atomic resonance, OD$_0$ the (resonant) peak optical density and $\Gamma$ the natural linewidth. In our system $\Gamma$ is power-broadened from $6$ to 8$\,$MHz. The light phase shift due to the BEC is given via the Kramers-Kronig relation, i.e.\ $\delta\phi=-\textrm{OD}\,\Delta/\Gamma,$ with a maximum phase shift at $\Delta=\pm \Gamma/2.$ Under conditions similar to Fig.~\ref{fig:fig1}b, with $\textrm{OD}_0=0.4$ and $\Delta=0\,$MHz, in Fig.~\ref{fig:fig2} we investigate the theoretical fringe behaviour for ten BEC fringes with $\textrm{OD}_0=0.4$ and visibility $80\%$ observed with on-resonant imaging, as well as the key effect of changing the imaging beam detuning (Fig.~\ref{fig:fig2}).

By taking $z$ slices through the Talbot optical density `carpet' in Fig.~\ref{fig:fig2}b,e the visibility as a function of $x$ for imaging detuning $\Delta=0\,$MHz can be obtained, and compared to the behaviour for detunings $\Delta=\pm4\,$MHz (Fig.~\ref{fig:fig2} c,f). The model we use to fit fringes at a given $z$ is a sine wave of visibility $C$, modulated by a parabolic amplitude to incorporate the typical experimental behaviour of the fringes due to their origin from the overlap of two initially spatially separated clouds: 
\begin{equation}
(A_0+A_2(z-z_0)^2)(C \sin(k z+\phi)+1).\label{fitt}\end{equation} Visibilities $C>1$ occur when the absorption imaging beam shows regions of negative optical density -- where more light is in the beam after it passes through the BEC fringes than before -- a single-shot experimental example of which is shown in Fig.~\ref{fig:fig3}a with $C\approx150\%$. Fig.~\ref{fig:fig3}b presents the corresponding spatial Talbot experimental visibility data to Fig.~\ref{fig:fig2}e, at the same detunings, for imaging light interacting with periodic BEC fringes. For both positive and negative detuned light the experimental and theoretical visibility agree well and are enhanced from $~80\,$\% to $135\,$\% (Fig.~\ref{fig:fig3}b). This visibility enhancement for detuned light is due to the phase variation in the initial conditions converting reversibly into intensity variation upon propagation. The enhancement can also be interpreted as focusing since the sinusoidal spatial phase variation is like a collection of alternating positive and negative lenses. 

\begin{figure}[!b]
\includegraphics[width=0.95\columnwidth]{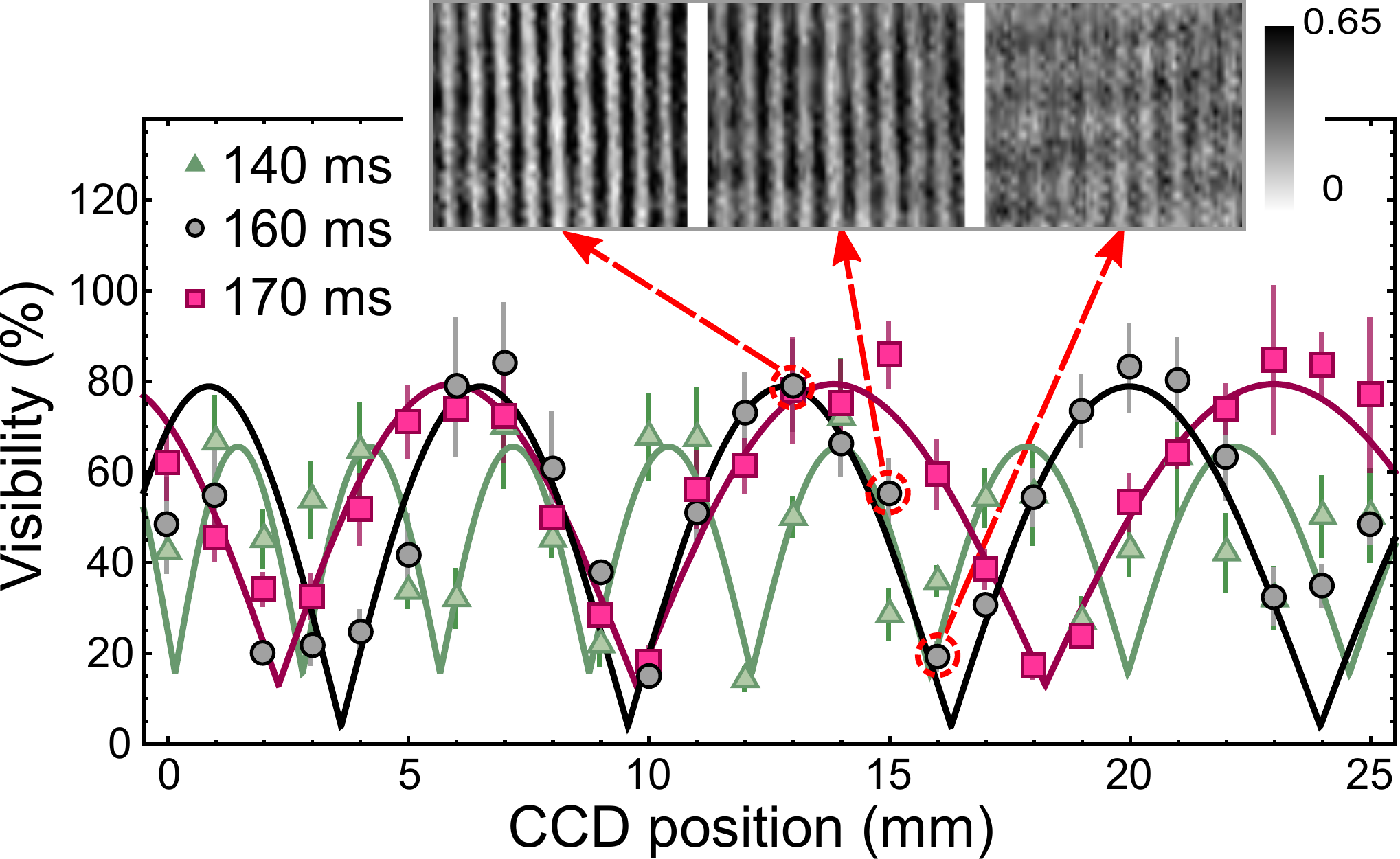}
\caption{Talbot period variation with fringe period. Visibility is observed as a function of camera position for $\Delta=+ 0.5\,$MHz imaging light after 140, 160 and $170\,$ms levitation. From fits to these data Talbot periods were measured to be $3.0(0.1)\,$mm, $6.1(0.2)\,$mm and $7.6(0.2)\,$mm, respectively, which agree with the theoretical Talbot periods of $3.4(0.1)\,$mm, $6.3(0.2)\,$mm and $8.1(0.8)\,$mm determined from the experimental fringe periods associated with the $140$, $160$ and $170\,$ms levitation times. Inset optical density plots ($320\times 280\,\mu$m$^2$) show the observed visibility changing through the Talbot period.}
\label{fig:4}
\end{figure} 

The variation of the observed fringe visibility as a function of the camera position can be attributed to the Talbot effect. The magnification of the imaging system modifies the Talbot period, which is therefore given by 
\begin{equation}
\label{Eq:TalbotPeriod}
\Lambda=\frac{(M\lambda_{f})^{2}}{\lambda}~,
\end{equation}
for light of wavelength $\lambda$ incident on a periodic density/phase modulation of period $\lambda_f$ where $M$ is the camera magnification. With the camera position at $13\,$mm, the observed fringes periods on the CCD for on-resonance light in Fig.~\ref{fig:fig3} were $70.9(1.5)\,\mu$m. The corresponding Talbot periods where measured to be $6.1(0.1)\,$mm, which is in good agreement with the theoretically predicted Talbot period of $6.4(0.3)\,$mm. 

To verify that the Talbot effect was at work we changed the period of the interference pattern, thus changing the Talbot period. The fringe period was altered by using different levitation times (Fig.~\ref{fig:fig1}c), $140$ and $170\,$ms were chosen as they would result in a notable change in the observed Talbot period. The fringe period under the experimental conditions in Fig.~\ref{fig:4}, with the camera at $13\,$mm and $\Delta=+0.5\,$MHz, were measured to be $51.2(0.3)\,\mu$m and $79.5(3.7)\,\mu$m for 140 and $170\,$ms respectively. The measured Talbot periods were found to be $3.0(0.1)\,$mm for $140\,$ms and $7.6(0.2)\,$mm for $170\,$ms, which are also in good agreement with the theoretical Talbot periods of $3.4(0.1)\,$mm and $8.1(0.8)\,$mm for 140 and $170\,$ms respectively. Fig.~\ref{fig:4} also presents density plots of the interference fringes at the peak, middle and trough of a Talbot period, showing the drastic effect the spatial Talbot effect can have on the interference signal.   

\section{optimal visibility}

\begin{figure}[!b]
\includegraphics[width=0.9\linewidth]{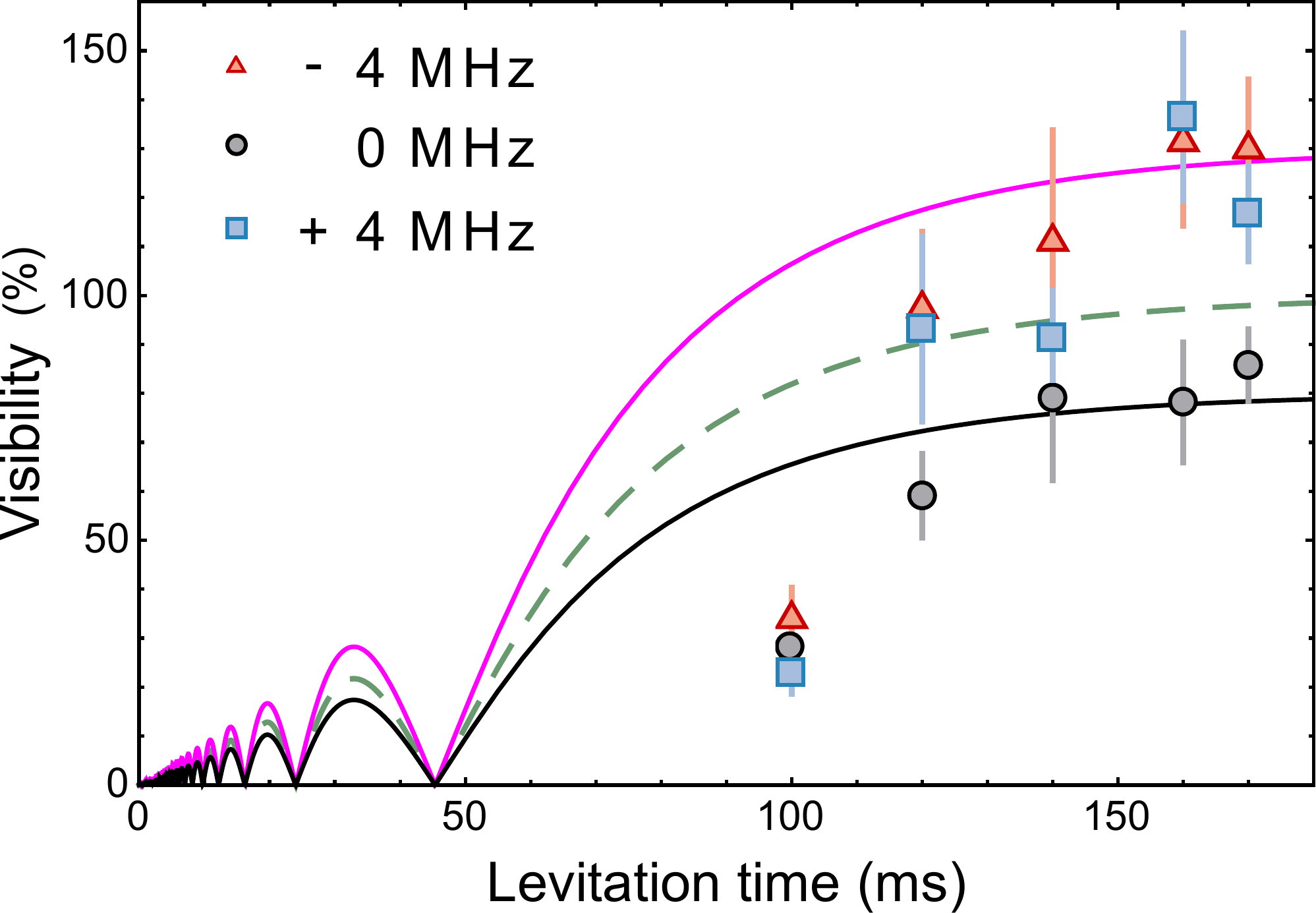}
\caption{Interference visibility as a function of levitation time and imaging detuning. Red circles, black squares and blue triangles denote detunings $\Delta= -4\,$MHz, $0\,$MHz and $+4\,$MHz, respectively. For comparison theory curves are provided of optimal visibility using Eq.~\ref{sinc} with pixellation-free final visibility values of $C=0.8,$ 1.0 and 1.3, shown by the black, green and cyan curves, respectively. The $C$ values 0.8 and 1.3 are chosen to approximately match the theoretical and experimental values from Figs.~\ref{fig:fig2} and \ref{fig:fig3}, where $\textrm{OD}_0=0.4$.}
\label{fig:fig5}
\end{figure}

In Fig.~\ref{fig:fig5} fringe visibility vs.\ expansion time is shown for both experiment and theory, with experimental data reaching the highest visibility reported in this type of interferometer. 
Note that for each detuning in Fig.~\ref{fig:fig5} the data at different levitation time were selected from the peak value within one Talbot period. For comparison to our experimental visibility vs. levitation time results shown in Fig.~\ref{fig:fig5}, we first consider the theoretical maximum visibility that can be reached. Assuming two perfect matter wave point sources, sinusoidal interference fringes result of the form $F(x)=1+C \sin(2\pi x/\lambda_f)$, where $\lambda_f$ is the fringe period, and $C$ is the visibility. When these fringes hit the $j^{\textrm{th}}$ pixel of a perfect CCD camera, with pixel size $l$, the fringe signal averaged over the range $(j-1/2)l \leq x \leq (j+1/2)l$ is 
\begin{equation} 
F_{\textrm{CCD}}=1+\textrm{sinc}(\pi l/\lambda_f) C \sin(2\pi j l/\lambda_f),
\label{sinc}\end{equation} 
i.e.\ the original fringe visibility $C$ reduces to $C'=\textrm{sinc}(\pi l/\lambda_f) C$. In our experiment $l=5\,\mu$m is constant, and the visibility only depends on the time-dependent fringe period -- which we know from the fit to Eq. \ref{eq:newfringeperiod} in Fig.~\ref{fig:fig1}c. This allows us to determine the theoretical maximum visibility sinc curves shown in Fig.~\ref{fig:fig5}, which for levitation times after 120$\,$ms reaches visibilities more than $0.95C$ and is surprisingly similar to experimental data. This is quite remarkable as in the theory we assumed perfect matter wave point sources without any allowance for our non-tomographic imaging which is susceptible to any spatial asymmetry in the source BECs and hence 3D curvature of the fringes (cf.\ Ref.~\cite{Interference1}). The reason for a resonant contrast $<100\%$ is attributed to either a slight angle between the fringe planes and the imaging beam $\textbf{k}$-vector, or population asymmetry in the split BECs -- however it is still the highest resonant visibility seen in this type of interferometer. 


\section{conclusions}

We report the first observation of the spatial Talbot effect of imaging light interacting with periodic BEC interference fringes. This interpretation is confirmed by the strong agreement between the Talbot lengths predicted by the fringe period and the measured Talbot periods from the experimental data. We have shown that the spatial Talbot effect can have drastic effects on the interference signal, and is therefore relevant for all BEC interferometers. More importantly, this effect can then be used as a tool, or diagnostic, to focus interference experiment imaging systems and help obtain the maximum possible visibility. By using detuned imaging light ($\pm4\,$MHz) the Talbot effect converts the imaging light periodic phase shift from the BEC fringes into an intensity ripple, which results in an increase in the observed fringe visibility. This enhancement in visibility allows us to observe single-shot interference visibility (as defined in Eq.~\ref{fitt}) of $C>135\,$\%, which is the highest ever reported in this type of interferometer. Such a contrast is not unphysical because the Talbot effect redistributes light into different parts of the absorption imaging beam -- leading to some beam regions with increased power, i.e.\ negative optical densities. The limit to attaining even higher visibility  is probably due to the underlying $80\%$ resonant visibility. This is either due to a slight angle between the imaging beam and the fringes, or population imbalance in the split BEC -- we have shown that it is unlikely to be due to CCD camera pixellation. 

Talbot-enhanced interferometers will be particularly advantageous in situations where the fringe period/camera pixel size ratio is small (or even less than 1), 
as the longer Talbot period can be used to accurately infer the matter wave fringe period. Moreover, by appropriately tailoring the incident imaging light to have variable focal distance (e.g.\ with a variable focus lens \cite{donner}), the entire 2D Talbot carpet (Fig.~\ref{fig:fig2}e) can be mapped out automatically. With the modification of varying imaging light propagation direction (e.g. with crossed AOMs \cite{excitations,arb1,arb2}) and a single-pixel detector \cite{padgett}, this would extend the scheme to short imaging wavelength detection and enable \textit{single shot} 2D Talbot mapping. Access to the full 2D Talbot carpet would enable more accurate phase determination of the fringes.

An advantage of our axial inverted parabolic levitation potential is that even in a compact vacuum system, with large initial condensate separation, it exponentially magnifies fringe periods to the same level seen in drop-tower \cite{microgravity} and $10\,$m vacuum experiments \cite{delta2}, however a caveat is that samples in the potential are also exponentially sensitive to initial position and velocity. In future we will enable accurate metrology with the interferometer by reducing our rms centre-of-mass noise. To further increase future interferometer phase accumulation time, the split BEC matter waves could be guided in a waveguide \cite{waveguide} or ring trap \cite{waveguide2,setup,otherrings2,Barry,UQ}, and delta kicking could be used to minimise atomic velocity spread \cite{delta2, delta, deltaopt2}.

The dataset for this paper is available online \cite{dataset}.

\section*{Funding Information}

The Leverhulme Trust (RPG-2013-074), DSTL (DSTLX-1000095638R), and EPSRC (EP/M013294/1).

\section*{Acknowledgments}

We greatly appreciate valuable discussions with T.\ Ackemann, G.-L.\ Oppo, G.\ R.\ M.\ Robb and W.\ J.\ Firth.

\end{document}